\documentclass[aps,prl,reprint,groupedaddress,footinbib,superscriptaddress, notitlepage,longbibliography]{revtex4-2}
\usepackage{amsmath}
\usepackage{graphicx}
\usepackage{hyperref}
\usepackage{bbold}
\usepackage{color}
\usepackage[normalem]{ulem}

\begin{document}
	\title{Superconducting diode effect in diffusive superconductors and Josephson junctions with Rashba spin-orbit coupling}
	\date{\today}
\author{Stefan Ili\'{c}}
\affiliation{Department of Physics and Nanoscience Center, University of Jyväskylä,
P.O. Box 35 (YFL), FI-40014 University of Jyväskylä, Finland}
\author{Pauli Virtanen}
\affiliation{Department of Physics and Nanoscience Center, University of Jyväskylä,
P.O. Box 35 (YFL), FI-40014 University of Jyväskylä, Finland}
\author{Daniel Crawford}
\affiliation{Department of Physics and Nanoscience Center, University of Jyväskylä,
P.O. Box 35 (YFL), FI-40014 University of Jyväskylä, Finland}
\author{Tero T. Heikkilä}
\affiliation{Department of Physics and Nanoscience Center, University of Jyväskylä,
P.O. Box 35 (YFL), FI-40014 University of Jyväskylä, Finland}
\author{F. Sebasti\'{a}n Bergeret}
\affiliation{Centro de F\'isica de Materiales (CFM-MPC) Centro Mixto CSIC-UPV/EHU, E-20018 Donostia-San Sebasti\'an,  Spain}
\affiliation{Donostia International Physics Center (DIPC), 20018 Donostia--San Sebasti\'an, Spain}
\begin{abstract}
We characterize the superconducting diode effect (SDE) in two-dimensional diffusive structures with Rashba spin-orbit coupling using the quasiclassical formalism. We consider both homogeneous superconductors and Josephson junctions. In the first case, the diode effect monotonically increases as the magnetic field is increased and the temperature is reduced, in contrast to the non-monotonic behavior found in clean structures. In Josephson junctions, SDE dramatically increases and changes its sign close to the $0-\pi$ transition of the junction, which occurs at specific junction lengths and strengths of the magnetic field. We show that the SDE is strongly suppressed in narrow junctions. Our results are relevant for understanding recent experiments that measure SDE in mesoscopic nanostructures, where significant disorder is unavoidable. 
 \end{abstract}
\maketitle
\emph{Introduction.-} The search for non-reciprocal transport in superconductors is a recent topic of major interest. The 
superconducting diode effect (SDE)\cite{nadeem2023superconducting} 
denotes the capacity of a superconducting system to have a direction-dependent critical current. The effect is closely related to magnetoelectric effects and can be observed when time reversal symmetry is broken in any superconducting structure with  gyrotropic symmetry\cite{kokkeler2024nonreciprocal}. Gyrotropy is present when inversion symmetry is broken, either on a microscopic scale (e.g., non-centrosymmetric crystals)\cite{agterberg2007magnetic,he2020magnetoelectric}, mesoscopic scales (e.g., lateral Josephson junctions)\cite{kokkeler2022field}, or macroscopic scales (e.g., asymmetric SQUIDs and films)\cite{souto2022josephson,fominov2022asymmetric,ciaccia2023gate,cuozzo2024microwave,hou2023ubiquitous}. For that reason, it is not surprising that the SDE has been observed in many different systems \cite{ando2020observation, baumgartner2022supercurrent, bauriedl2022supercurrent, wu2022field, jeon2022zero,  diez2023symmetry,hou2023ubiquitous}.

Beyond its technological interest, the SDE is also intriguing because it might be a macroscopic transport manifestation of microscopic properties of the materials, particularly the spin-orbit coupling (SOC) strength. Indeed, the SDE has been mainly  studied in mesoscopic systems with Rashba SOC \cite{yokoyama_anomalous_2014, he2022phenomenological, yuan2022supercurrent, daido_intrinsic_2022, ilic2022theory, baumgartner2022supercurrent}. Most theoretical works focus on the ballistic limit, and only  few studies have considered the effect of disorder\cite{ilic2022theory,kokkeler2022field}. 

\begin{figure}[h!]
    \centering
    \includegraphics[width=\columnwidth]{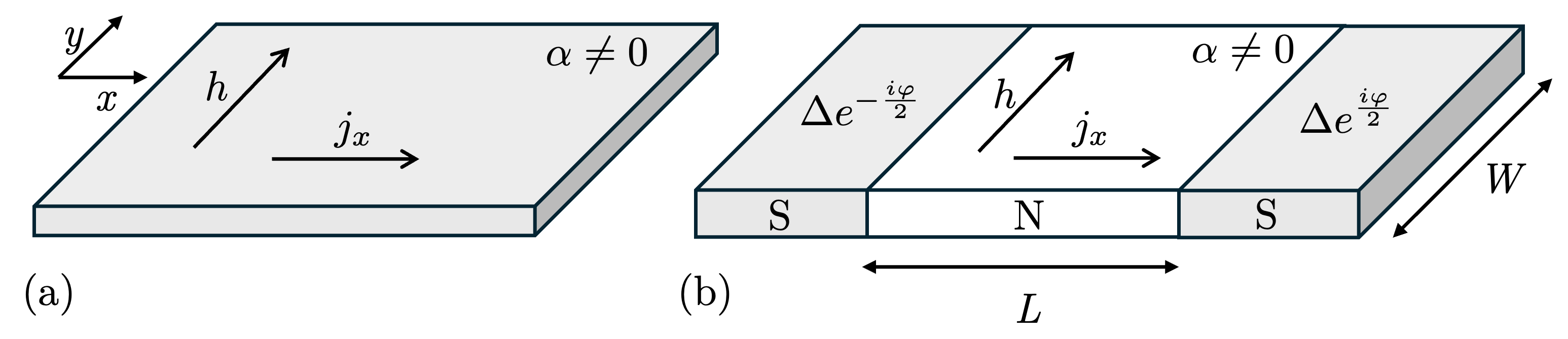}
    \caption{Schematic representation of setups considered in this work. (a) Bulk superconductor. (b) Josephson S/N/S junction, with length $L$ and width $W$.} 
    \label{figSet}
\end{figure}

In this work, we present a comprehensive theoretical analysis of the SDE in a diffusive superconducting system with Rashba spin-orbit coupling (SOC). We focus on isotropic Rashba superconductors [shown in Fig.~\ref{figSet}(a)] as well as Josephson junctions composed of a Rashba medium  sandwiched between two superconductors [shown in Fig.~\ref{figSet}(b)]. Our analysis uses the recently derived Usadel equation for superconducting systems with intrinsic SOC\cite{virtanen2022nonlinear}. In Rashba superconductors the diode effect monotonically increases as the magnetic field is increased and temperature is reduced. In Josephson junctions the SDE is very large close to the $0$-$\pi$ transition where it also changes its sign. We also show how  the SDE  depends on the transverse dimensions of the Josephson junction. In particular our theory predicts a reduction of the effect when the width is reduced.

Before starting the full calculation of the SDE, it is instructive to explain the effect in general terms and introduce some notations.
Breaking both inversion and time-reversal symmetries in bulk superconductors and Josephson junctions leads to the inequalities
\begin{equation}
j(q) \neq -j(-q), \qquad J(\varphi) \neq -J(-\varphi),
\label{Eq:ineq}
\end{equation}
 where $j(q)$ is the bulk supercurrent, $q=\nabla \varphi$ is the superconducting phase gradient,  $J(\varphi)$ is the Josephson current, and $\varphi$ is the phase difference across the Josephson junction. These inequalities are at the base of the non-reciprocal  phenomena. First, they allow for an appearance of anomalous currents, namely $j(0)\neq 0$ and $J(0)\neq 0$ \cite{edelstein1995magnetoelectric, yip2002two, edelstein2005magnetoelectric}. Second, the ground state (with zero current) of an infinite bulk system is associated with a finite phase gradient $j(q_0)=0$ --- this is known as helical superconductivity \cite{agterberg2007magnetic, dimitrova2007theory, houzet2015quasiclassical}. Analogously, in a Josephson junction the current-phase relation (CPR) acquires an anomalous phase shift $J(\varphi_0)=0$, which is known as the anomalous Josephson effect  or the $\varphi_0$ effect \cite{buzdin_direct_2008, bergeret_theory_2015, konschelle_theory_2015}. Another consequence of Eq.~\eqref{Eq:ineq} is that the critical currents in the two directions may be unequal, namely $j_c^+ \neq -j_c^-$ and $J_c^+\neq - J_c^-$, where $j_c^+=\text{max}[j(q)]$, $j_c^-=\text{min}[j(q)]$, $J_c^+=\text{max}[J(\varphi)]$ and $J_c^-=\text{min}[J(\varphi)]$, and $\pm$ denotes the direction of the current flow. This is the SDE: in the current window between $j_c^+$ and $j_c^-$ in the bulk, or between $J_c^+$ and $J_c^-$ in junctions, the supercurrent can flow in one direction, whereas only a dissipative normal current can flow in the other. Next, we introduce the main equations we use to quantify the SDE.

\emph{Quasiclassical equations.-} The Usadel equation for systems with any linear-in-momentum intrinsic SOC has the general structure \cite{virtanen2022nonlinear}
\begin{equation}
\tilde{\partial}_i \check{\mathcal{J}}_i+[(\omega_n-i h_i\sigma_i)\tau_z+\check{\Delta},\check{g}]=0.
\label{Eq:Usadel}
\end{equation}
Here, $\check{g}$ is the quasiclassical Green's function, which is a matrix in spin-Nambu space, and which satisfies the normalization condition $\check{g}^2=1$. The spin and Nambu space are spanned by Pauli matrices $\sigma_i$ and $\tau_i$ $(i=x,y,z)$, respectively. $\omega_n=2\pi T (n+\frac{1}{2})$, $n\in Z$, is the Matsubara frequency, $T$ is the temperature and $h_i$ is the Zeeman or exchange field acting in $i$-direction. Superconductivity is accounted for by $\check{\Delta}=i\Delta \tau_+-i\Delta^* \tau_-$, where $\Delta$ is the superconducting order parameter and $\tau_\pm=\frac{1}{2}(\tau_x\pm i\tau_y)$. The covariant derivative $\tilde{\partial}_i=\partial_i-i[\mathcal{A}_i,\cdot]$ $(i=x,y,z)$ contains the vector potential $\mathcal{A}_i=\mathcal{A}_i^j \sigma_j$, which captures the effect of any linear-in-momentum SOC. The matrix current is $\check{\mathcal{J}}_i=\check{\mathcal{J}}_i^{0}+\check{\mathcal{J}}_i^{H}$, with
\begin{align}
\check{\mathcal{J}}_i^0&=-D \check{g} \tilde{\partial}_i \check{g} \nonumber \\
\check{\mathcal{J}}_i^H&=-\frac{D\tau}{4m} [\{\mathcal{F}_{ij}+\check{g} \mathcal{F}_{ij}\check{g}, \tilde{\partial}_j \check{g}\}-i \tilde{\partial}_j (\check{g}[\tilde{\partial}_i \check{g}, \tilde{\partial}_j \check{g}])].
\end{align}
Here, $\check{\mathcal{J}}_i^0$ is the usual diffusive contribution,  $D=\frac{1}{2} v_F^2 \tau$ is the diffusion constant in a two dimensional system, $v_F$ is the Fermi speed in the absence of SOC, and $\tau$ is the elastic scattering time due to disorder. The second term, $\check{\mathcal{J}}_i^H$, captures all magnetoelectric phenomena, such as spin-galvanic effects \cite{edelstein1995magnetoelectric, yip2002two, edelstein2005magnetoelectric} and non-reciprocal phenomena \cite{agterberg2007magnetic,dimitrova2007theory,buzdin_direct_2008,bergeret_theory_2015}, including the diode effect. Here, $\mathcal{F}_{ij}=-i [\mathcal{A}_i, \mathcal{A}_j]$. Equation \eqref{Eq:Usadel} is valid within the quasiclassical and diffusive approximations, meaning that $E_F\gg \tau^{-1}\gg \Delta, h, \mathcal{A}_i p_F$, where $E_F$ and $p_F$ are the Fermi energy and momentum. The order parameter $\Delta$ satisfies the self-consistency condition 
\begin{equation}
\Delta \ln \frac{T}{T_c}= 2\pi T \sum_{\omega_n>0} \left(
-\frac{i}{2} \text{Tr} [\tau_+ \check{g}]-\frac{\Delta}{\omega_n}
\right),
\label{Eq:SelfCons}
\end{equation}
where $T_c$ is the critical temperature. 

In the following, we  focus on the specific case of Rashba SOC, where $\mathcal{A}_x=m\alpha \sigma_y$, $\mathcal{A}_y=-m\alpha \sigma_x$, and $\mathcal{F}_{xy}=-\mathcal{F}_{yx}=2m^2\alpha^2\sigma_z$. We assume that the field is applied along the $y$-direction, $h\sigma_y$, and consider supercurrent flow along the $x$-direction (see Fig.~\ref{figSet}). The Usadel equation can be significantly simplified if the considered system is infinite in the $y$-direction, which is a good approximation for large isotropic systems or wide Josephson junctions. In this case, the structure of the Green's function is simply $\check{g}=\check{g}_0+\check{g}_t \sigma_y$, where the first and second term correspond to the singlet and triplet components, respectively. Moreover, system properties vary only along the $x$-direction, while they are constant along the $y$-direction, so the Usadel equation becomes effectively a 1D problem.  Next, we introduce $\check{g}_\pm=\check{g}_0 \pm \check{g}_t$, and the Usadel equation becomes
\begin{multline}
0=[(\omega_n\mp ih)\tau_z +\check{\Delta}, \check{g}_\pm]- D (\check{g}_\pm \check{g}_\pm ')'\mp \frac{\Gamma_r}{4}[\check{g}_+,\check{g}_-]\\
\pm \frac{\xi \Gamma_{st}}{4} \left( \{\check{g}_\pm ', \{\check{g}_+,\check{g}_-\} \}+(\check{g}_+-\check{g}_-) \check{g}_\mp' (\check{g}_+-\check{g}_-) \right),
\label{Eq:UsadelPM}
\end{multline}
with the normalization condition $\check{g}_\pm^2=1$. Here, we introduced the notation $A'=\partial_x A$,  and the coherence length $\xi=\sqrt{D/\Delta_0}$, where $\Delta_0$ is the order parameter at $h=0$ and $T=0$. The effect of SOC is captured by two energy scales: the Dyakonov-Perel relaxation rate $\Gamma_r=2 \alpha^2 p_F^2 \tau$, and the singlet-triplet conversion rate $\Gamma_{st}=\Gamma_r^{3/2} \Delta_0 ^{1/2}/(2E_F)$. The former describes triplet relaxation, while the latter is the driving force of the diode effect (and other magnetoelectric effects). We see that necessarily $\Gamma_r \gg \Gamma_{st}$, since $E_F\gg \Delta, \Gamma_r$. The supercurrent along the $x$-direction is $j_x=-\frac{i\pi \nu T}{2} \sum_{\omega_n>0} \text{Tr} [\tau_z \check{\mathcal{J}}_x]$, which simplifies to
\begin{multline}
j_x=\frac{i\pi \nu T}{2} \text{Tr} \tau_z \sum_{\omega_n>0, \pm} \left[
D \check{g}_\pm \check{g}_{\pm}'
+\frac{\xi \Gamma_{st}}{4} \check{g}_\pm [\check{g}_+,\check{g}_-]\right].
\label{Eq:CurrentPM}
\end{multline}

To numerically solve the Usadel equation \eqref{Eq:UsadelPM} we employ the Riccati parametrization, namely 
 \begin{equation}
 \check{g}_\pm= \frac{1}{(1+\gamma_\pm \tilde{\gamma}_\pm)}
 [(1-\gamma_\pm \tilde{\gamma}_\pm)\tau_z+2 \gamma_\pm \tau_+ + 2\tilde{\gamma}_\pm \tau_-],
 \end{equation}
 which automatically satisfies the normalization condition. The explicit form of Eq.~\eqref{Eq:UsadelPM} parameterized this way is given in the Supplementary Information (SI). In the following we explore the SDE in 2D superconductors and Josephson junctions.

\begin{figure}[h!]
    \centering
    \includegraphics[width=\columnwidth]{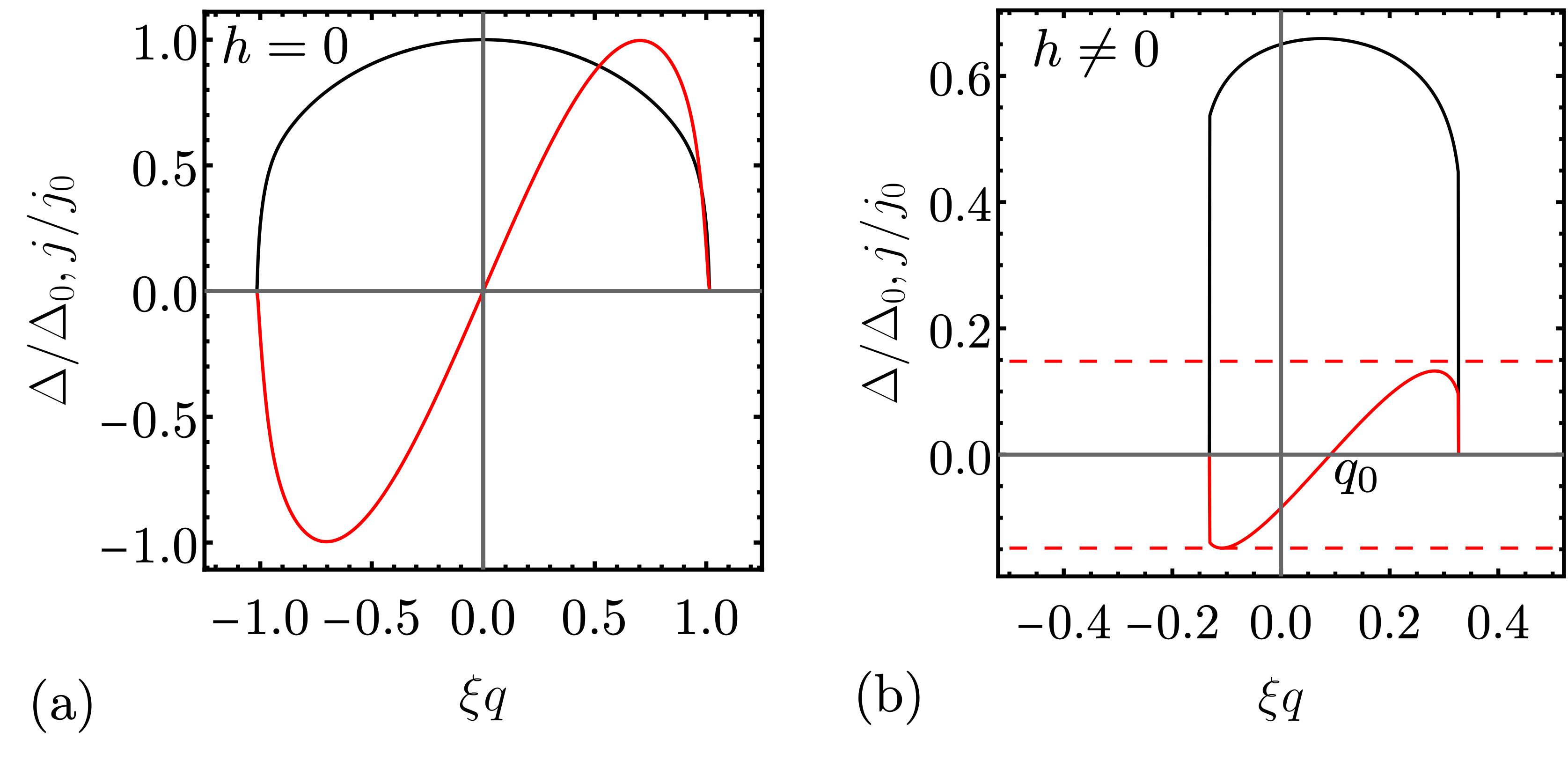}
    \caption{Order parameter $\Delta$ (black curves), and the supercurrent $j$ (red curve) calculated self-consistently as a function of the phase gradient $q$ at (a) $h=0$  and (b) $h=1.3 \Delta_0$. Current is normalized with $j_0$, which is the critical current at $h=0$ and $T=0$. The dashed red lines in (b), corresponding to the current $j=\pm j_c^{-}$, serve to illustrate that $j_c^+\neq -j_c^-$ ($\eta \approx 5\%$). Parameters used in the plots: $\Gamma_r=5 \Delta_0$, $\Gamma_{st}=0.25 \Delta_0$ and $T=0.05 T_c$.} 
    \label{figBulk}
\end{figure}

\emph{SDE in a two dimensional superconductor with SOC.-} We first consider an isotropic 2D  system, depicted in Fig.~\ref{figSet}(a). Here, we may take $\Delta(x)=\Delta e^{i q x}$, where $q=q_0+\delta q$ is the phase gradient (also called Cooper pair momentum) with two contributions: the intrinsic modulation $q_0$ created by the interplay of the Zeeman field and SOC, and $\delta q$ that comes from an externally applied supercurrent. Then, the  Riccati functions can be expressed as $\gamma_\pm (x)=i \rho_\pm e^{i q x}$ and $\tilde{\gamma}_\pm (x)=-i \rho_{\pm} e^{- i q x}$, and  the Usadel equation \eqref{Eq:UsadelPM} reduces to algebraic equations for $\rho_\pm$ written in the SI. Using the solution of these equations together with Eqs.~\eqref{Eq:SelfCons} and \eqref{Eq:CurrentPM}, we find $\Delta$ and $j_x$ self-consistently as a function of the phase gradient $q$. Importantly, the self-consistency condition can yield multiple solutions for $\Delta$ in the vicinity of the first-order phase transition \cite{heikkila2019thermal}, which happens at low temperatures and high fields if the SOC is not too strong ($\Gamma_r \lesssim 10 \Delta_0$). In those cases, we choose the solution that minimizes the free energy (see the SI). The helical modulation $q_0$ can be determined from the condition $j_x(q_0)=0$. Some examples of this calculation at zero and finite magnetic field \footnote{Note that the field chosen in Fig.~\ref{figBulk}(b), $h=1.3 \Delta_0$, is above the standard Clogston-Chrandrasekhar limit \cite{buzdin2005proximity} $h_{c2}=\Delta_0/\sqrt{2}$. Superconductivity can still persist at such a strong field due to the presence of SOC, which greatly enhances the upper critical field. } are shown in Fig.~\ref{figBulk}. At finite field depicted in panel (b), we see a finite $q_0$, and more importantly a small difference in critical currents  $j_c^+ \neq - j_c^-$. 

\begin{figure}[h!]
    \centering
    \includegraphics[width=\columnwidth]{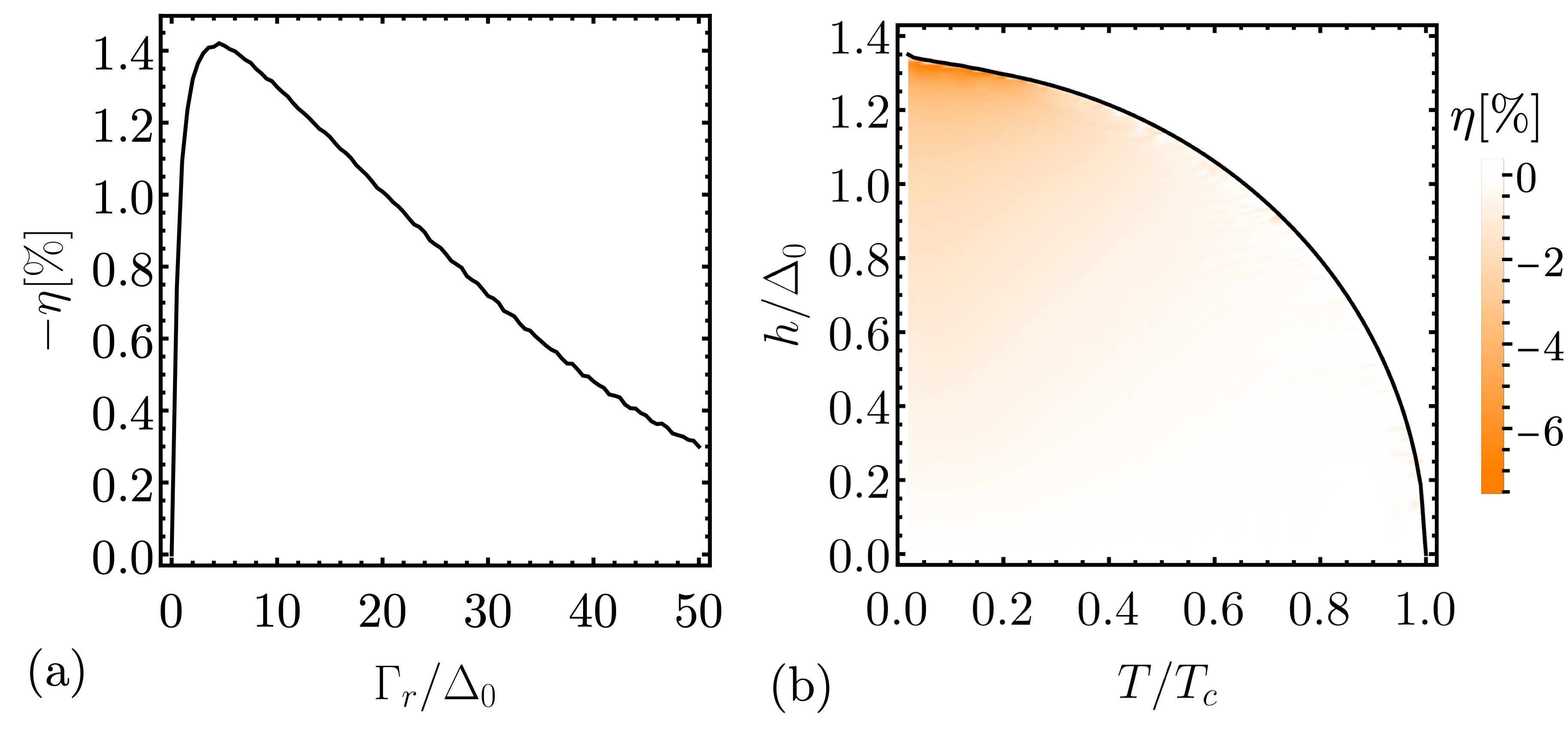}
    \caption{SDE in bulk Rashba superconductors. (a) Efficiency $\eta$ as a function of the spin-relaxation rate $\Gamma_r$, taking $\Gamma_{st}=0.025 \Gamma_r^{3/2}/\sqrt{\Delta_0}$, $h=0.7 \Delta_0$ and $T=0.05 T_c$. (b) Efficiency $\eta$ calculated at every point of the $h-T$ phase diagram. We set $\Gamma_r=5\Delta_0$ and $\Gamma_{st}=0.25 \Delta_0$.} 
    \label{figEtaBulk}
\end{figure}

We proceed to investigate the diode efficiency defined as
\begin{equation}
\eta=(j_c^+-|j_c^-|)/(j_c^++|j_c^-|).
\end{equation}
In Fig.~\ref{figEtaBulk}(a), we show how $\eta$ depends on the strength of SOC at some fixed field and temperature. Weak SOC naturally yields a weak diode effect, whereas too strong SOC suppresses the diode effect due to the rapid triplet relaxation. An optimal strength of SOC is when  $\Gamma_r$ is of the order of a few $\Delta_0$. Then $\Gamma_{st}\propto \Gamma_r^{3/2}\Delta^{1/2}/E_F$ is sufficiently large to create a substantial SDE, without having too strong relaxation that would suppress the effect. In the SI, we demonstrate analytically that the SDE vanishes for $\Gamma_r\gg \Delta_0, h$, even though $q_0$ may be finite. 

In Fig.~\ref{figEtaBulk}(b), we calculate $\eta$ for every point in the $h-T$ phase diagram. We see that, in contrast to the clean case \cite{daido_intrinsic_2022,ilic2022theory}, $\eta$ is monotonous as a function of $h$ and $T$, without sign changes. This comes from the fact that helical superconductivity in the clean case has a more complex behavior, determined by the interplay of the two helical bands \cite{agterberg2007magnetic, dimitrova2007theory}, which contribute differently to helical superconductivity at low and high fields. Importantly, this interplay is absent in the diffusive case, where the helical bands are always strongly coupled by disorder, and consequently this leads to a simpler behavior of the SDE.  The findings shown in \ref{figEtaBulk}(b) are similar to Ref.~\onlinecite{ilic2022theory}, which studied $\eta$ in the vicinity of the superconducting phase transition in a different parameter regime ($E_F \gtrsim \alpha p_F \gg \tau^{-1}$). However, here we manage to explore the SDE in the full temperature range.

\begin{figure*}[t!]
    \centering
    \includegraphics[width=\textwidth]{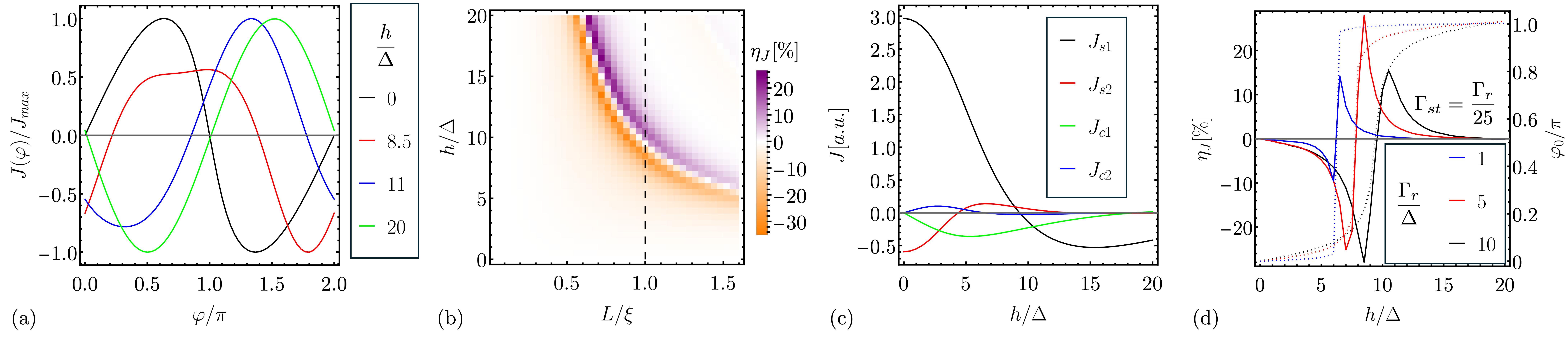}
    \caption{SDE in wide Josephson junctions. (a) CPR at various strengths of the magnetic field $h$. We normalize each CPR by the maximal value of the current $J_{max}=\text{max}|J(\varphi)|$. (b) Efficiency $\eta_J$ as a function of magnetic field and junction length. The dashed line corresponds to $L=\xi$, used in panels (a), (c) and (d).  (c) First two sine and cosine harmonics of the CPR as a function of magnetic field at $L=\xi$. (d) Diode efficiency $\eta$ (full lines) and the anomalous phase shift $\varphi_0$ (dotted lines) as a function of magnetic field, for different values of $\Gamma_r$ and $\Gamma_{st}$, at $L=\xi$ and $T=0.1 \Delta$. Parameters used in (a), (b) and (c): $T=0.1 \Delta$, $\Gamma_r=10 \Delta$, $\Gamma_{st}=0.4 \Delta$.} 
    \label{figetaJJ}
\end{figure*}

\emph{SDE in wide Josephson junctions.-} Let us now turn to SDE in Josephson junctions with a large width $W\gg \xi$ [see Fig.~\ref{figSet}(b)]. In this case,  we may again use the one-dimensional Usadel equation given in Eq.~\eqref{Eq:UsadelPM}. We consider a S/N/S junction where S are superconducting electrodes, while N is a normal medium with Rashba SOC and a Zeeman field. We solve Eq.~\eqref{Eq:UsadelPM} in Riccati parametrization  assuming perfectly transparent interfaces between S and N, namely $\check{g}_\pm (0)=\check{g}_l$ and $\check{g}_\pm (L)=\check{g}_r$. Here,  $g_{l,r}=(\omega_n \tau_z- i \Delta \tau_y e^{\pm i\varphi \tau_z/2})/\sqrt{\omega_n^2+\Delta^2}$ are the Green's functions of the superconducting leads, $L$ is the length of the junction, and $\varphi$ is the phase difference across the junction (see details in the SI). Combining this solution with the expression for the current \eqref{Eq:CurrentPM}, we obtain the current-phase relation (CPR) $J(\varphi)$. Some examples of CPR are shown in Fig.~\ref{figetaJJ}(a) for different values of the  field $h$. We see that CPR exhibits different phase shifts $\varphi_0$ as well as the difference in critical currents $J_c^+\neq -J_c^-$.

The  CPR can be expanded in harmonics as $J(\varphi)=\sum_{n=1}^{\infty} [J_{sn} \sin (n\varphi)+J_{cn} \cos(n\varphi)]$. In diffusive junctions, the lowest-order harmonics ($n=1,2$) give the dominant contribution to the CPR \cite{heikkila2002supercurrent}. The cosine harmonics can only exist when both time-reversal and inversion symmetries are broken, and therefore they are responsible for the SDE. In order to see the SDE, it is essential that the CPR contains prominent second-order harmonics, which may only occur in junctions with high interface transparency. The SDE is strongest when the first and second harmonics are comparable in magnitude.

We next introduce the efficiency of a Josephson diode similarly as in the bulk:
\begin{equation}
\eta_J=(J_c^+-|J_c^-|)/(J_c^++|J_c^-|).
\end{equation}
We plot $\eta_J$ as a function of $h$ and junction length $L$ in Fig.~\ref{figetaJJ}(b). We see that the diode effect is strongly enhanced and changes its sign at some specific values of $h$ and $L$. To better understand this, we  plot the first two sine and cosine harmonics of the CPR in Fig.~\ref{figetaJJ}(c) as a function of $h$. SDE is strongest at fields where the dominant first harmonic $J_{s1}$ is close to its sign change. This corresponds to the $0-\pi$ transition studied in S/F/S junctions, because the CPR of the junctions changes from $J(\varphi)\approx |J_{c1}| \sin\varphi$ at low fields to $J(\varphi)\approx |J_{c1}| \sin(\varphi+\pi)$ at high fields. Note, however, that in contrast to conventional S/F/S structures, this transition is smoothed in our case by the phase shift $\varphi_0$, which can take any value between 0 and $\pi$ by changing $h$. The $0-\pi$ transition can also be achieved by tuning $L$ with fixed $h$. We see that $J_{s1}$ dominates the CPR away from the $0-\pi$ transition, but close to the transition other harmonics become comparable, which yields a large diode effect. The sign change of $\eta$ comes from the sign change of $J_{s1}$ at the  transition. Similar enhancement of SDE accompanied by the sign change has been predicted in Josephson junctions with twisted bilayer graphene close to the $0-\pi$ transition \cite{hu2023josephson}, and in ballistic multichannel junctions with Rashba SOC \cite{costa2023sign} resulting from the interplay of different anomalous shifts of separate channels.

In Fig.~\ref{figetaJJ}(d) we illustrate how the diode efficiency $\eta$ correlates with the anomalous phase shift $\varphi_0$. The latter is defined as the ground state of the junction, obtained by minimizing the Josephson energy $E_J(\varphi)\propto \int_0^\varphi J(\chi) d\chi$. At low fields, both $\varphi_0$ and $\eta$ scale roughly linearly with $h$, but this is no longer the case at higher fields. 

Comparing these results to the bulk case, we see that the junction geometry allows for significantly stronger SDE than in a bulk superconductor. This comes mostly from the fact that in the junction we may impose fairly strong magnetic fields $h$, while in the bulk the field strength is limited by the paramagnetic limit \cite{buzdin2005proximity}. Such strong fields in junctions may be experimentally achieved by for example  ferromagnetic insulators, which can induce a strong exchange field in the $N$ part of the junction \cite{moodera1988electron,strambini2017revealing,zhang2020phase}.

Ref.~\onlinecite{costa2023sign} studied the SDE in a 2DEG heterostructure subjected to an in-plane Zeeman field. They report a $\eta(h)$ with a very similar shape to Fig.~\ref{figetaJJ} - there is a strong enhancement of SDE accompanied by a sign-reversal at some field. This finding was well described using a ballistic multichannel short junction model. Our results suggest an alternative explanation -- the same features in $\eta(h)$ appear also in diffusive and moderately long junctions ($L\sim \xi$).

\emph{Finite-size effects in Josephson junctions.-}
In narrow junctions with finite width $W \lesssim \xi$, the simplified Usadel equation ~\eqref{Eq:UsadelPM} no longer holds. Namely, in narrow junctions the SOC induces a circulating supercurrent flow \cite{bergeret2020theory}, the system properties are inhomogeneous along both $x$ and $y$ directions, and one has to deal with the full 2D problem. Moreover, in finite-width samples one needs to consider all three triplet components of the Green's function, $\check{g}=\check{g}_0+\check{g}_t^i \sigma_i$, due to the triplet precession created by the SOC.  Therefore, in order to study finite-size effects in Josephson junctions, the general Usadel equation Eq.~\eqref{Eq:Usadel} needs to be numerically solved. We assume that the $N$ part of junction is in the region $(x,y)\in [0,L]\times [0,W]$. Again, we  assume transparent interfaces between $N$ and $S$, $\check{g}(0,y)=\check{g}_l$ and $\check{g}(L,y)=\check{g}_r$.  At the sample edges along the transversal direction we impose vacuum interfaces, such that the matrix current vanishes: $\check{\mathcal{J}}_{y}(x,0)=\check{\mathcal{J}}_{y}(x,W)=0$. We finally calculate the supercurrent in $x$ and $y$ directions using $j_i=-\frac{i\pi \nu T}{2} \sum_{\omega_n>0} \text{Tr} [\tau_z \check{\mathcal{J}}_i].$  Additional details about the solution procedure along with numeric codes are presented elsewhere \cite{virtanen2024}. In the following we discuss the consequences of finite $W$ on the diode effect. 

\begin{figure}[h!]
    \centering
    \includegraphics[width=0.9 \columnwidth]{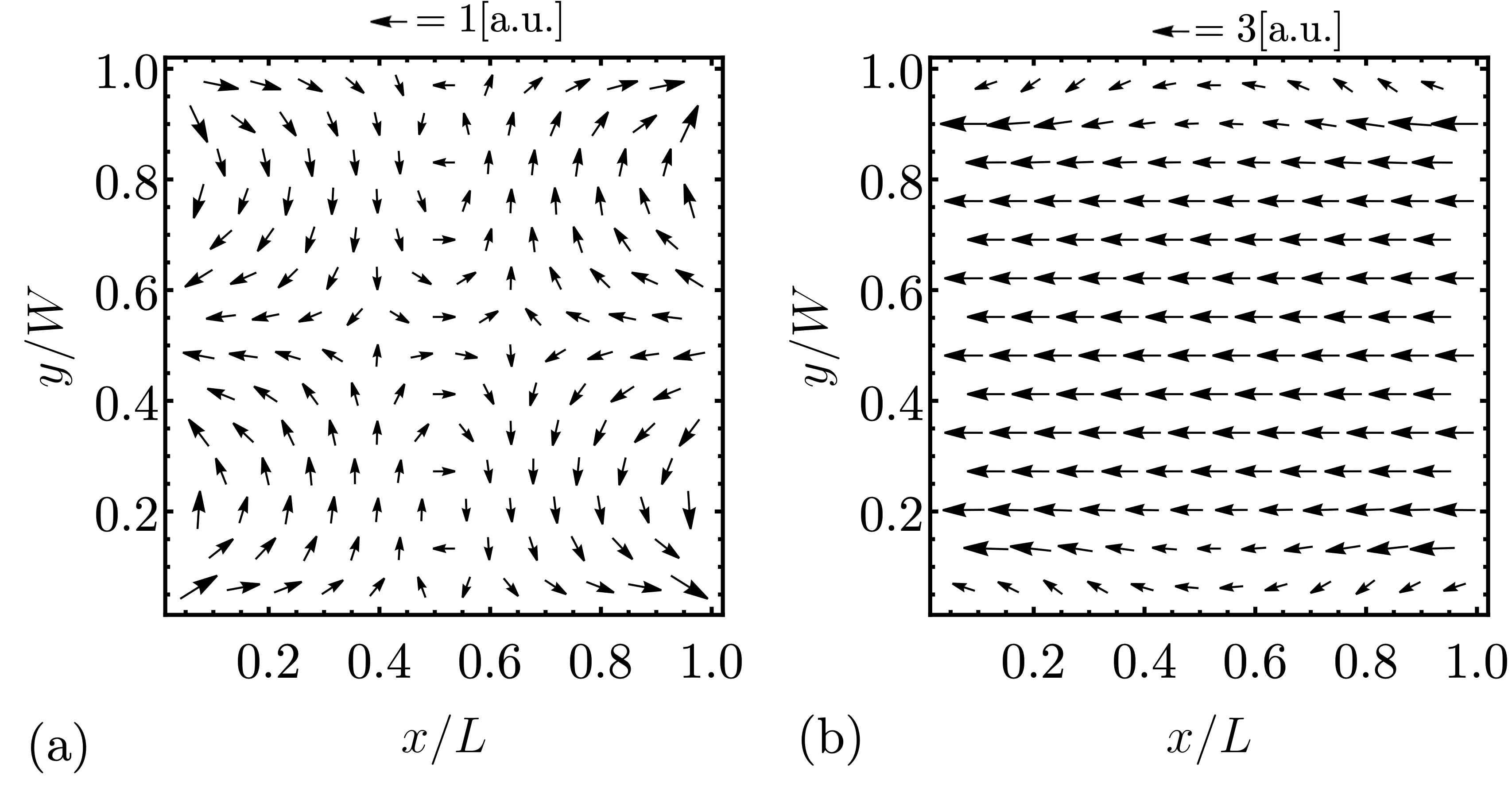}
    \caption{Distribution  of the anomalous current ($\varphi=0$) in a (a) narrow junction with $L=1.2 \xi$, $W=0.5 \xi$, and (b) wide junction $L=1.2\xi$, $W=4\xi$.
    Parameters used: $T=0.1 \Delta$, $h=10\Delta$, $\Gamma_r=10 \Delta$, $\Gamma_{st}=0.4 \Delta$. Note that there is different scaling of arrows in (a) and (b), as indicated above the figures.} 
    \label{figJ}
\end{figure}

\begin{figure}[h!]
    \centering
    \includegraphics[width=\columnwidth]{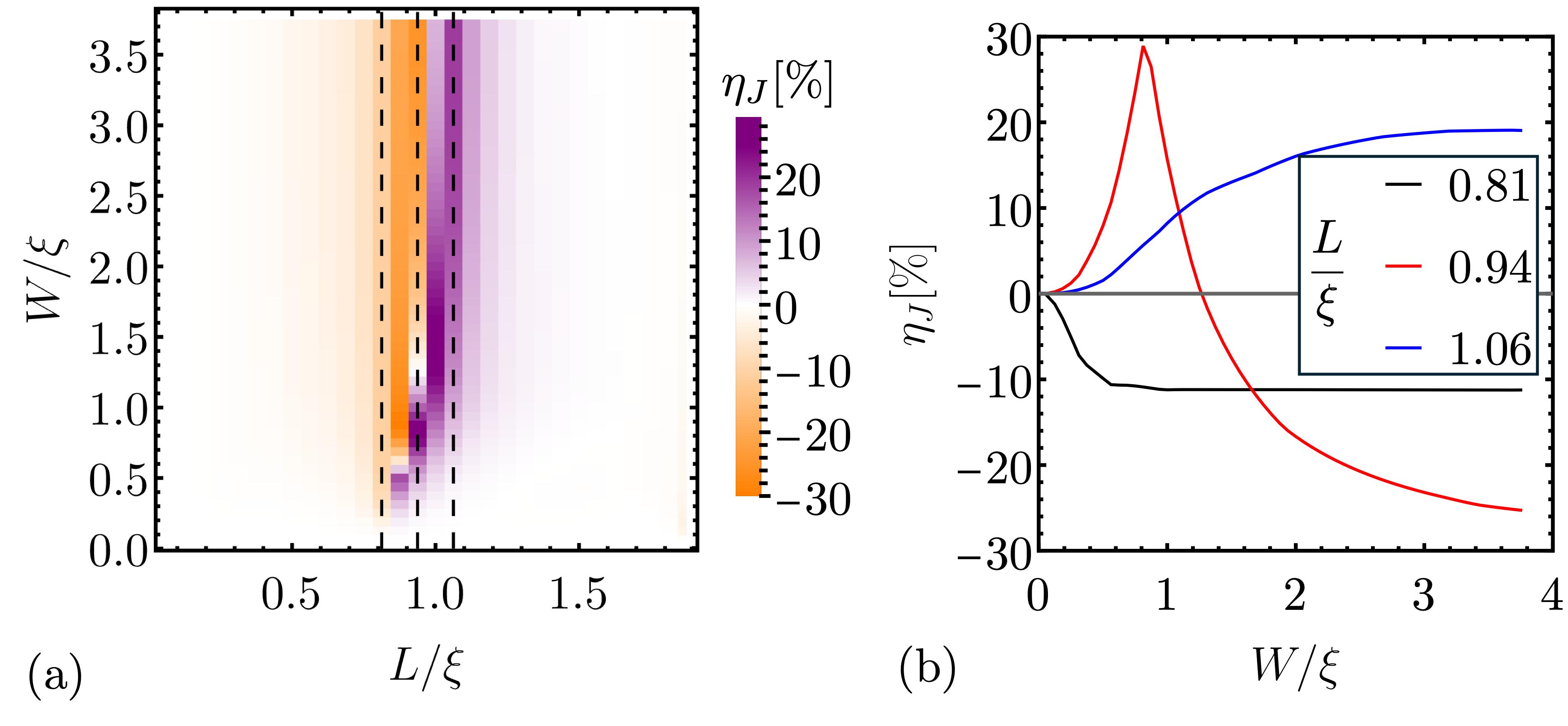}
    \caption{SDE in a finite-sized Josephson junction. (a) Efficiency $\eta_J$ as a function of junction length $L$ and width $W$. Dashed lines correspond to curves in (b). (b) $\eta_J$ as a function of $W$ for several different $L$.  We use the same parameters as in Fig.~\ref{figJ}} 
    \label{figLW}
\end{figure}

In Fig.~\ref{figJ} we show the distribution of the anomalous current $(\varphi=0)$ at a finite magnetic field. In a narrow junction shown in panel (a), we see that the flow is highly inhomogeneous, with SOC-induced circulating currents. As $W$ is increased, the current becomes uniform, as shown in panel (b). In Fig.~\ref{figLW}, we plot the diode efficiency as a function of $W$ and $L$. We see that $\eta_J$ is strongly suppressed in very narrow junctions ($W\ll \xi$). This is because in this limit the system behaves effectively like a 1D Rashba nanowire \cite{ojanen_magnetoelectric_2012,baumard2020interplay}, where the magnetic field $h_y$ alone is not sufficient to produce magnetoelectric phenomena (an additional field component applied along $x$ would be needed).   
Diode efficiency grows as $W$ is increased, and finally saturates to the wide-junction value. As illustrated in Fig.~\ref{figLW}(b), $\eta_J$ may change its sign as a function of $W$ if the junction is tuned close to the $0-\pi$ transition.

\emph{Summary.-} In summary, we present a theory of the superconducting diode effect in diffusive structures with Rashba SOC, both in bulk systems and Josephson junctions. We do so by exploiting the recently established quasiclassical Usadel equation for systems with intrinsic SOC~\cite{virtanen2022nonlinear}. In the bulk, we characterize the SDE in the entire $h-T$ phase diagram.  In Josephson junctions, we show that the SDE is strongest close to the $0-\pi$ transition of the junction, and that the effect is suppressed in very narrow junctions.  The same approach may be used to study diode effects in the presence of arbitrary  linear-in-momentum SOC (e.g. Dresselhaus type), for different junction geometries (e.g. lateral junctions), and for different interfaces described by an appropriate boundary condition.

\begin{acknowledgements}
This work was supported by the Research Council of Finland (Contract No.~355056 and 354735)
and European Union's HORIZON-RIA programme (Grant Agreement No.~101135240 JOGATE). F.S.B. acknowledges financial support from Spanish MCIN/AEI/
10.13039/501100011033 through projects PID2020-114252GB-I00 (SPIRIT)
and TED2021-130292B-C42,  the Basque Government through grant IT-1591-22, and
European Union’s Horizon Europe research and innovation programme under grant agreement No 101130224 (JOSEPHINE). We acknowledge grants of computer capacity from the Finnish Grid and
   Cloud Infrastructure (persistent identifier
   urn:nbn:fi:research-infras-2016072533 ).
\end{acknowledgements}

\onecolumngrid
\pagebreak
\clearpage

\setcounter{equation}{0}
\setcounter{figure}{0}
\setcounter{table}{0}
\setcounter{page}{1}
\renewcommand{\thefigure}{S\arabic{figure}}
\renewcommand{\theequation}{S\arabic{equation}}
\begin{center}
\textbf{\large Supplemental Material for ``Superconducting diode effect in diffusive superconductors and Josephson junctions with Rashba spin-orbit coupling"} \\
\end{center}
This Supplemental Material contains technical details and derivations omitted in the main text. 

In Sec.~I, we explicitly write Eq.~(5) from the main text in Riccati parametrization. We also provide a simplified version of these expressions valid for homogeneous superconductors. We produce the majority of results in the main text by solving these equations, as shown in Figs.~(2)-(4) of the main text.  In Sec.~II, we write the free energy  of a diffusive Rashba superconductors, which is in some cases needed to determine the stable solution of the self-consistency condition,  Eq.~(4) of the main text. This is, in particular, relevant for the discussion of SDE in homogeneous superconductors presented in the main text. In Sec.~III, we derive the Usadel equation in the limit of strong spin relaxation $\Gamma_r\gg \Delta_0, h$. The result resembles the equations previously derived for diffusive Rashba superconductors in another parameter regime ~\cite{houzet2015quasiclassical} ($E_F \gtrsim \alpha p_F \gg \tau^{-1}\gg \Delta_0$). Based on the form of this result, it is evident that the SDE vanishes in this parameter regime.  
\section{I. Riccati parametrization \label{App1}}
The Green's function $\check{g}_{\pm}$ has the following form in Riccati parametrization
\begin{equation}
\check{g}_\pm=\frac{1}{1+\gamma_\pm \tilde{\gamma}_\pm} 
\begin{pmatrix}
1-\gamma_\pm \tilde{\gamma}_\pm & 2 \gamma_\pm \\
2 \tilde{\gamma}_\pm & -1+\gamma_\pm \tilde{\gamma}_\pm
\end{pmatrix}  =
\begin{pmatrix}
G_\pm & F_\pm \\
\tilde{F}_\pm & -G_\pm
\end{pmatrix}.
\end{equation}
The Usadel equation, Eq.~(5) of the main text, now becomes
\begin{multline}
D\gamma_\pm''=-i\Delta -i \Delta^* \gamma_\pm^2+2 (\omega_n\mp ih)\gamma_\pm + D \tilde{F}_{\pm} (\gamma_\pm')^2  -\frac{\Gamma_r}{4} F_\mp K_\pm (1+\gamma_\pm \tilde{\gamma}_{\mp}) \\
\pm \frac{\xi \Gamma_{st}}{4} F^2_\mp K_\pm^2 \tilde{\gamma}_\mp' 
\pm \xi \Gamma_{st} (G_+ G_- + \frac{1}{2} F_+\tilde{F}_-+\frac{1}{2} \tilde{F}_+F_- ) \gamma_\pm',
\label{Eq:Riccatti1}
\end{multline}
\begin{multline}
D\tilde{\gamma}_\pm''=i \Delta^* + i \Delta \tilde{\gamma}_\pm^2+2 (\omega_n\mp ih)\tilde{\gamma}_\pm + D F_{\pm} (\tilde{\gamma}_\pm')^2-\frac{\Gamma_r}{4} \tilde{F}_\mp \tilde{K}_\pm (1+\tilde{\gamma}_\pm \gamma_{\mp})
\\
\mp \frac{\xi \Gamma_{st}}{4} \tilde{F}^2_\mp \tilde{K}_\pm^2 \gamma_\mp'\mp 
\xi \Gamma_{st} (G_+ G_- + \frac{1}{2} F_+\tilde{F}_-+\frac{1}{2} \tilde{F}_+F_- ) \tilde{\gamma}_\pm',
\label{Eq:Riccatti2}
\end{multline}
where we introduced $K_\pm=1-\gamma_\pm/\gamma_{\mp}$, $\tilde{K}_\pm=1-\tilde{\gamma}_\pm/\tilde{\gamma}_{\mp}$.
Next, the supercurrent given by Eq.~(6) of the main text reduces to
\begin{equation}
j_x=2 i \pi \nu T \sum_{\omega_n>0} \left( 
D \frac{\gamma_+ \tilde{\gamma}_+'-\gamma_+' \tilde{\gamma}_+}{(1+\gamma_+\tilde{\gamma}_+)^2}
+
D \frac{\gamma_- \tilde{\gamma}_-'-\gamma_-' \tilde{\gamma}_-}{(1+\gamma_-\tilde{\gamma}_-)^2}
+\Gamma_{st}\xi \frac{(1+\gamma_+ \tilde{\gamma}_-)(1+\gamma_-\tilde{\gamma}_+)(\gamma_+\tilde{\gamma}_+-\gamma_-\tilde{\gamma}_-)}{(1+\gamma_+\tilde{\gamma}_+)^2 (1+\gamma_-\tilde{\gamma}_-)^2}
\right).
\label{Eq:RiccattiJ}
\end{equation}
Finally, the self-consistency condition, Eq.~(4) of the main text, becomes
\begin{equation}
\Delta \ln \frac{T}{T_c}= 2\pi T \sum_{\omega_n>0} \left(
-\frac{i}{2}(F_++F_-)-\frac{\Delta}{\omega_n}
\right).
\label{Eq:RiccattiDelta}
\end{equation}
It can be used to determine the suppression of the order parameter from its value at zero temperature, zero exchange field and vanishing phase gradient. The critical temperature, exchange field or current are then determined by finding their values corresponding to the case where $\Delta=0$ is the only solution of the self-consistency equation.

\subsection{A. Homogeneous superconductors}
In homogeneous superconductors we may take $\Delta(x)=\Delta e^{i q x}$, with $\gamma_\pm (x)=i \rho_\pm e^{i q x}$ and $\tilde{\gamma}_\pm (x)=-i \rho_{\pm} e^{- i q x}$. Then, Eqs.~\eqref{Eq:Riccatti1} and \eqref{Eq:Riccatti2} reduce to algebraic equations for $\rho_\pm$:
\begin{multline}
(2\omega_n\mp ih) \rho_\pm +D q^2 \rho_\pm \frac{1-\rho_\pm^2}{1+\rho_\pm^2}\pm \Gamma_r \frac{(\rho_+-\rho_-)(1+\rho_+\rho_-)}{1+\rho_\mp^2}
\\
\pm q\xi \Gamma_{st} \left( 
\frac{\rho_\mp (\rho_+-\rho_-)^2}{(1+\rho_\mp^2)^2}
-\rho_\pm \frac{(1-\rho_-^2)(1-\rho_+^2)+4 \rho_+\rho_-}{(1+\rho_+^2)(1+\rho_-^2)}
\right)
=(1-\rho_\pm^2)\Delta. 
\end{multline}
The current \eqref{Eq:RiccattiJ} becomes
\begin{equation}
j_x= -i\pi \nu T \sum_{\omega_n>0} \left[
4 i D q \left( \frac{\rho_+^2}{1+\rho_+^2}+\frac{\rho_-^2}{1+\rho_-^2}\right)+
\frac{2 \Gamma_{st} \xi (\rho_-^2-\rho_+^2)(1+\rho_+\rho_-)^2}{(1+\rho_+^2)^2 (1+\rho_-^2)^2}
\right],
\end{equation}
and the self-consistency condition \eqref{Eq:RiccattiDelta} yields
\begin{equation}
\Delta \ln \frac{T}{T_c}=2\pi T\sum_{\omega_n>0} \left(
\frac{\rho_+}{1+\rho_+^2}+
\frac{\rho_-}{1+\rho_-^2}
-\frac{\Delta}{\omega_n}
\right).
\label{eq:riccattiselfcons}
\end{equation} 
Typically below the critical temperature, exchange field or current there are two solutions: the stable solution with a finite $\Delta$ corresponding to the superconducting state and the unstable solution with $\Delta=0$. Above the critical temperature, field and current the normal-state solution $\Delta=0$ becomes stable. However, in the case of a non-vanishing exchange field, there is a parameter region with more than two solutions of the free energy, with multiple (meta)stable solutions. In those cases we need to evaluate the free energy corresponding to those solutions to determine the most stable one.

\section{II. Free energy \label{App2}}
The free energy of a diffusive superconductors with arbitrary linear-in-momentum SOC can be readily derived using the non-linear sigma model established in Ref.~\onlinecite{virtanen2022nonlinear}. It reads  
\begin{equation}
F_S=\frac{\nu \Delta^2}{V}-\frac{\pi T \nu}{2} \sum_{\omega_n}\text{Tr} 
\left[
(\omega_n-i h_i \sigma_i)\tau_z \check{g}+\check{\Delta} \check{g}-\frac{1}{4} D (\tilde{\partial}_i \check{g})^2+ \frac{D  \tau}{4 m} \mathcal{F}_{ij} \check{g} \tilde{\partial}_i \check{g} \tilde{\partial}_j \check{g}
\right] 
\end{equation}
where $V$ is the pairing constant. One can check that the self-consistency condition, Eq.~(4) of the main text, is obtained by finding the extrema of $F_S$ with respect to $\Delta$, i.e., setting $\partial_\Delta F_S=0$. Using the self-consistency condition in the form $\frac{\Delta^2}{V}=\frac{\pi T}{4}\sum_{\omega_n} \text{Tr}\check{\Delta}\check{g}$, we can express the free energy as
\begin{equation}
F_S=-\frac{\pi T \nu}{2} \sum_{\omega_n}\text{Tr} 
\left[
(\omega_n-i h_i \sigma_i)\tau_z \check{g}+\frac{1}{2}\check{\Delta} \check{g}-\frac{1}{4} D (\tilde{\partial}_i \check{g})^2+ \frac{D  \tau}{4 m} \mathcal{F}_{ij} \check{g} \tilde{\partial}_i \check{g} \tilde{\partial}_j \check{g}
\right]. 
\end{equation}

In the following, we focus on homogeneous Rashba superconductors, with magnetic field applied along the $y$-direction and the current flowing along the $x$-direction, as considered in the main text. The free energy difference of the superconducting and normal state written in terms of bulk Riccati functions $\rho_\pm$ is then
\begin{multline}
\delta F=F_S-F_N=\pi \nu T \sum_{\omega_n} \bigg[
\sum_\pm \frac{1}{1+\rho_\pm^2}\left(  2\omega \rho_\pm^2 -\Delta \rho_\pm \pm 2 i h + \frac{D q^2 \rho_\pm^2}{1+\rho_\pm^2}\right)
\\
+
\frac{\Gamma_r (\rho_+-\rho_-)^2}{2(1+\rho_+^2)(1+\rho_-^2)}
+\frac{i q \xi \Gamma_{st} (\rho_+^2-\rho_-^2)(1+\rho_+\rho_-)}{(1+\rho_+^2)^2(1+\rho_-^2)^2}
\bigg].
\label{Eq:RiccattiFree}
\end{multline}
The above expression is used to choose the most stable solution of the self-consistency condition in cases when it yields multiple non-zero solutions for $\Delta$. This happens only at low temperatures and high magnetic fields, close to the first-order phase transition to the normal state \cite{buzdin2005proximity}. Increasing SOC suppresses the capacity of the superconductor to have a first-order phase transition and multiple nonzero solutions of $\Delta$. At $\Gamma_r\gtrsim 10 \Delta_0$ the phase transition to the normal state is always of the second order. For the specific parameters chosen in Fig.~3(b) of the main text ($\Gamma_r=5\Delta_0$, $\Gamma_{st}=0.25 \Delta_0$), only a very small part of the phase diagram at high fields ($h>1.3 \Delta_0$) shows multistability of $\Delta$. 

\section{III. Limit of strong triplet relaxation \label{App3}}
Let us first express the Usadel equation, Eq.~(5) of the main text, in terms of the singlet and triplet Green's functions, $\check{g}_0=\frac{1}{2} (\check{g}_++\check{g}_-)$ and $\check{g}_t=\frac{1}{2} (\check{g}_+-\check{g}_-)$. We obtain
\begin{equation}
[\omega_n \tau_z+\check{\Delta}, \check{g}_0]- i h [\tau_z, \check{g}_t]-D  (\check{g}_0 \check{g}_0'+\check{g}_t \check{g}_t')' 
\\+\xi \Gamma_{st} (\check{g}_0^2 \check{g}_t)'=0,  
\end{equation} 
\begin{equation}
[\omega_n \tau_z+\check{\Delta}, \check{g}_t] - i h[\tau_z, \check{g}_0]- D (\check{g}_0 \check{g}_t'+\check{g}_t \check{g}_0')' \\ +\Gamma_r \check{g}_0 \check{g}_t +\xi \Gamma_{st} \left(
\check{g}_0' \check{g}_0^2+ \check{g}_t [\check{g}_0', \check{g}_t]
\right)=0,
\end{equation}
with the normalization condition $\check{g}_0^2+\check{g}_t^2=1$ and $\{\check{g}_0,\check{g}_t\}=0$. 
We take the limit of strong spin relaxation $\Gamma_r \gg \Delta, h$. Under this condition, the triplet component is small $\check{g}_t \ll \check{g}_0$. The normalization condition becomes $\check{g}_0^2 \approx 1$ with $\{\check{g}_0, \check{g}_t\}=0$. Then, the Usadel equation simplifies to
\begin{equation}
[\omega_n \tau_z+\check{\Delta}, \check{g}_0]- i h [\tau_z, \check{g}_t]-D  (\check{g}_0 \check{g}_0')'
+\xi \Gamma_{st} \check{g}_t'=0,  
\label{Eq:S14}
\end{equation}
\begin{equation}
- i h[\tau_z, \check{g}_0]+\Gamma_r \check{g}_0 \check{g}_t+\xi \Gamma_{st} \check{g}_0' =0.
\label{Eq:S15}
\end{equation}
From Eq.~\eqref{Eq:S15} we get 
\begin{equation}
\check{g}_t=\frac{i h }{\Gamma_r} \check{g}_0  [\tau_z, \check{g}_0]-\frac{\xi \Gamma_{st}}{\Gamma_r} \check{g}_0  \check{g}_0'.
\end{equation}
Substituting this into Eq.~\eqref{Eq:S14} gives
\begin{equation}
\left[\omega_n \tau_z+\check{\Delta}+\frac{\Gamma}{2} \tau_z \check{g}_0 \tau_z, \check{g}_0 \right]-D \partial_x (\check{g}_0 (\partial_x+\hat{\Lambda}) \check{g}_0) \\-D\hat{\Lambda} (\check{g}_0 \partial_x \check{g}_0)=0,
\label{eq:s14}
\end{equation}
where we introduced the operator  $\hat{\Lambda}=-\frac{i q_0}{2}[\tau_z,\cdot]$, with
\begin{equation}
\Gamma= \frac{2 h^2}{\Gamma_r}, \qquad q_0=\frac{4 \alpha h}{v_F^2}.
\end{equation}
Next, we notice that  Eq.~\eqref{eq:s14} can be further simplified if we add to the left-hand side the term $-D \hat{\Lambda} (\check{g}_0 \hat{\Lambda} \check{g}_0)$. This term scales as $q_0^2\sim \alpha^2/v_F^2$ and therefore it is negligible within the quasiclassical approximation (which assumes $v_F\gg \alpha$). Then, the Usadel equation becomes 
\begin{equation}
\left[\omega_n \tau_z+\check{\Delta}+\frac{\Gamma}{2} \tau_z \check{g}_0 \tau_z, \check{g}_0 \right]-D \tilde{\nabla}_x (\check{g}_0 \tilde{\nabla}_x \check{g}_0)=0,
\end{equation}
where we introduced $\tilde{\nabla}_x=\partial_x+\hat{\Lambda}=\partial_x-\frac{i q_0}{2} [\tau_z,\cdot]$. Finally, the current along the $x$-direction is
\begin{equation}
j_x=i \pi \nu T \, \text{Tr} \, \tau_z \sum_{\omega_n>0} D \check{g}_0 \tilde{\nabla}_x \check{g}_0.
\end{equation}
As we can see, the only effects of SOC and Zeeman field are a simple shift of the phase gradient by $q_0$, and a depairing effect captured by $\Gamma$. In other words, in this limit there is helical superconductivity with a modulation vector $q_0$, but not a diode effect. Similar equation was found by Houzet and Meyer \cite{houzet2015quasiclassical} in a different regime $\alpha p_F \gg \tau^{-1}\gg \Delta_0,h$. 
\end{document}